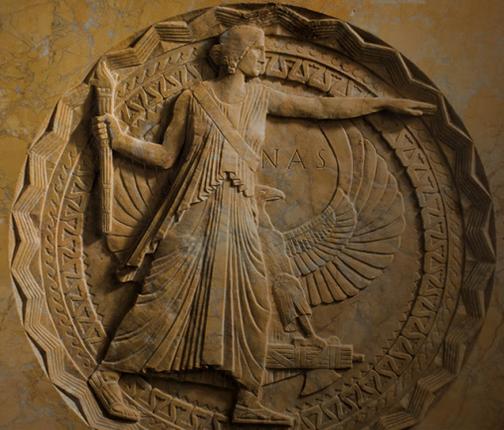

# BIOGRAPHICAL MEMOIRS

## JAMES BURKETT HARTLE

August 20, 1939–May 17, 2023
Elected to the NAS, 1991

*A Biographical Memoir by Gary T. Horowitz and Kip S. Thorne*

JAMES BURKETT HARTLE was a theoretical physicist who made major contributions to our understanding of relativistic stars, black holes, and cosmology. Most of his career, however, was devoted to studying the universe as a quantum system. As a result, he was known as the father of quantum cosmology. He is best known for two seminal papers with Stephen Hawking that introduced two quantum states of fundamental importance: the "Hartle-Hawking vacuum" for matter fields outside a black hole, and the "no-boundary wave function of the universe" for cosmology. Jim (as everyone called him) was a warm and caring person who was genuinely concerned with the success of his students, postdocs, and colleagues. He was generous with his time and helped to foster a culture of a welcoming family among gravitational physicists.

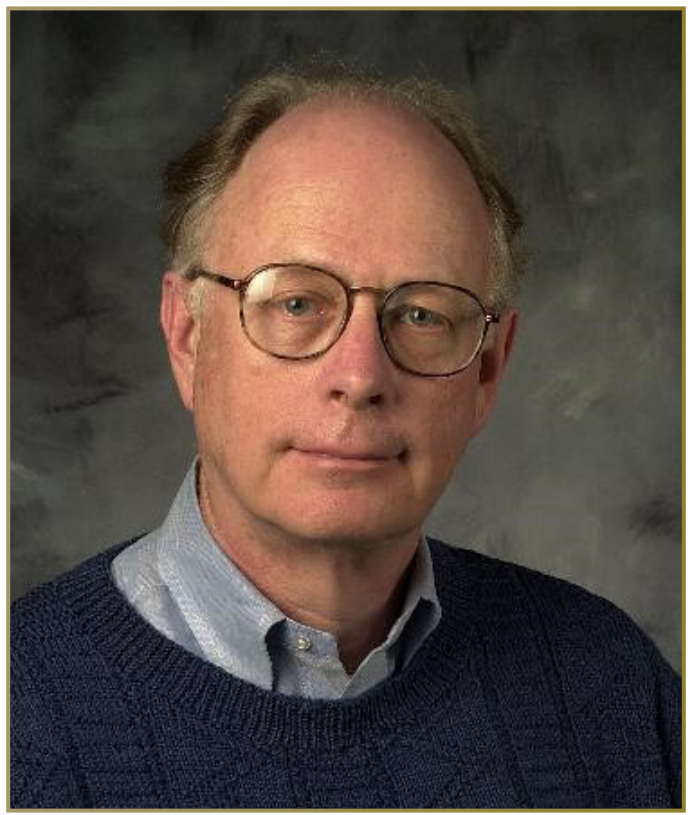

Figure 1  James Burkett Hartle.

### EARLY LIFE AND EDUCATION

Jim was born in Baltimore, Maryland, on August 20, 1939, to Charles James Hartle and Anna Elizabeth Burkett Hartle. He began his education at the Gilman School (a private all-boys school in Baltimore) but also attended public schools in Ohio and Connecticut because of family relocations. The family eventually returned to Baltimore, and he completed high school at the Gilman School in 1956. Jim's father was an executive at IBM and wanted his son to become an engineer, so that fall Jim entered Princeton University's School of Engineering.

In his freshman year, Jim took an introductory physics class taught (as luck would have it) by John Wheeler. Jim soon realized that he was much more interested in physics than engineering and changed his major. Over the next few years, he kept asking Wheeler for advice, and Wheeler became a lifelong mentor. For Jim's 1960 senior thesis, Wheeler suggested that he work with Wheeler's postdoc, Dieter Brill. Jim learned general relativity, and he and Brill together developed a way to describe weak gravitational waves in curved spacetime, including the waves' nonlinear interaction with each other. They used their formalism to model a spherical gravitational-wave geon: an object, envisioned by Wheeler, made from a bundle of gravitational waves that holds itself together by the waves' gravitational pull. Their remarkably

NATIONAL ACADEMY OF SCIENCES

powerful analysis, published four years later,[1] became the principal foundation for Richard Isaacson's 1968 solution to the most famous unsolved relativity problem of that era: what, in all generality, is the energy-momentum tensor of a gravitational wave and how does it act back on the curved spacetime through which the wave travels to alter that spacetime curvature.[2] Asked a decade later whether he had foreseen anything as an undergraduate along the lines of Isaacson's follow-on breakthrough, he responded, "I was not thinking in this direction at all. I just had deep satisfaction when I saw the geon hold itself together."

While at Princeton, Jim came across Richard Feynman's path integral formulation of quantum mechanics and was immediately struck by its simplicity and beauty. It became a cornerstone of his later approach to physics. After receiving his bachelor's degree from Princeton in 1960, Jim decided to go to the California Institute of Technology (Caltech) for graduate school, largely because Feynman was there. He was a graduate student assistant during Feynman's famous Lectures on Physics and was responsible for taping the lectures, photographing the blackboards, and several other tasks.[3] He also attended Feynman's lectures on gravitation in which Feynman basically derived general relativity starting from general principles.[4] Despite this, Jim ended up working with Murray Gell-Mann and finished his Ph.D. in only three years in the fall of 1963. Jim later described his experience as Gell-Mann's student as follows:[5] "To work with Murray it helps first of all not to have too fragile an ego. Murray was very accessible to his students. But I remember one occasion when I went to his office and said, Professor Gell-Mann, I have a question.' He replied somewhat irritably, 'Questions, questions—you are always coming in here with questions! Why don't you try bringing in some answers sometime!'" Even so, Jim got along well with Gell-Mann, and he became another lifelong mentor.

After receiving his Ph.D., Jim spent a year (1963–64) at the Princeton Institute for Advanced Study, during which he and Brill wrote their gravitational-wave geon paper[6] based on his 1960 senior thesis. Then for two years (1964–66) he was a physics instructor (and researcher) at Princeton University, before joining the faculty at the University of California, Santa Barbara (UCSB) in 1966. Aside from a two-year leave of absence at the University of Chicago from 1981 to 1983, during which he toyed with moving to Chicago permanently, Jim spent his entire post-Princeton career at UCSB.

## RESEARCH LIFE

Jim's research career was broadly divided up into three phases: relativistic astrophysics, quantum properties of black holes, and quantum cosmology. We now briefly review his main contributions in each of these areas.

## RELATIVISTIC ASTROPHYSICS

Prior to 1963, general relativity was a backwater in physics research—more the province of mathematicians than physicists. This changed rather quickly after Maarten Schmidt at Caltech discovered quasars, just a few months before Jim's departure for Princeton. Wheeler at Princeton and others elsewhere argued that a quasar's enormous power might come from the huge gravitational energy released in the collapse of a large mass to form a black hole. At Caltech, by contrast, Willy Fowler and Fred Hoyle argued that supermassive stars might power the quasars. When Fowler and Hoyle announced this in a Caltech seminar that Jim likely attended, Feynman objected that general relativity would make all supermassive stars unstable to gravitational collapse if they were not spinning, a result he had derived out of personal curiosity but never published.[7] A consensus quickly arose among the world's astrophysicists that general relativity was key to understanding quasars, and the field of relativistic astrophysics was born.

Jim Bardeen, Jim's officemate at Caltech, quickly gravitated into the orbit of Feynman and Fowler and wound up doing a Ph.D. thesis on the stability and pulsations of relativistic stars. At this time (1963), Kip Thorne, who had become friends with Bardeen and Hartle during his senior year at Caltech, was at Princeton and started working with Wheeler on the same topic. By 1965, Bardeen and Thorne were beginning to collaborate on relativistic stars and their pulsations, and Jim Hartle and David Sharp (at Princeton) were formulating a variational principle for the structure of rotating relativistic stars.[8] In 1966, Hartle moved to Santa Barbara, Thorne moved to Caltech, and Bardeen was in transition from Caltech to the University of Washington in Seattle, via a postdoc at the University of California, Berkeley. With the three permanently ensconced on the west coast of the United States, there arose a rich and fruitful interaction among them and their students and postdocs.

One of Jim's most important contributions to relativistic astrophysics was his work on slowly rotating relativistic stars. Between 1967 and 1975, he wrote a series of nine technical papers that laid out the general relativistic theory of slowly rotating, relativistic stars and explored applications that helped astrophysicists develop a strong physical understanding of these objects.[9–17] In his first paper he explained his motivations, which were all formulated before there was any firm observational evidence for the existence of these stars (either supermassive or neutron): first, Fowler's arguments that a supermassive star's rotation will, if large enough, stabilize it against gravitational collapse, so it can be long-lived by burning nuclear fuel; second, Wheeler's arguments that the star's rotation will couple its radial pulsational modes to its quadrupolar modes, enabling the radial modes to be damped by radiating gravitational waves; and third, Wheeler's arguments that the rotation of a neutron star might be the power source for supernova remnants such as the Crab nebula.

Jim explored each of these motivations in one or more of his nine papers. But in the half century since he wrote them, his papers' importance has continued to increase for these reasons: (a) the large set of observed astrophysical venues in which rotating relativistic stars have been found—pulsars, binary pulsars, X-ray binaries, gamma-ray bursters, kilonovae, magnetars, and gravitational-wave sources; (b) the fact that these objects' stars almost always rotate slowly enough for Jim's formalism to be valid to better accuracy than the observations; and (c) the fact that the observed stars' rotation almost always strongly influences the observed phenomena. Correspondingly, research on all these objects has relied heavily on theoretical modeling via Jim's formalism and on data analysis tools that depend on his formalism for their development.

In the 1970s, when Jim was doing this research, astronomers were beginning to identify candidate black holes in X-ray binaries and, from their observational data, estimating the mass of the compact X-ray emitting object. This mass was the best way to determine whether the compact object was a black hole or a neutron star. To make that determination reliably, they needed to know the maximum mass of a neutron star. That maximum mass was uncertain because of large uncertainties in the equation of state of matter above nuclear densities, $\rho_n \simeq 5 \times 10^{14}$ g cm$^{-3}$, so it was important to know an upper bound on the neutron-star maximum mass based on what was known reliably about the equation of state. Several researchers, including Hartle and his student, Augusto Gianni Sabbadini, deduced such upper bounds for nonrotating neutron stars. In 1978, Hartle published the most thorough and definitive analysis of this problem and a review of all efforts on it.[18] Additionally, he analyzed the influence of rotation on the maximum mass.[19,20]

In the near half century since then, as the observational venues for rotating neutron stars has burgeoned (binary pulsars, gamma-ray bursters, kilonovae, magnetars, gravitational-wave sources), it has become possible to begin extracting, from the observations, information about the equation of state (EOS) between one- and ten-times nuclear density. Jim's analyses of rotating relativistic stars have been key to this extraction. Particularly powerful for this in the coming few years are observations of a quantity on which Jim focused a lot of attention in his 1978 paper[21] and later: neutron-star moments of inertia ($I=J/\Omega$, where $J$ is the star's angular momentum, measured via its relativistic dragging of inertial frames, and $\Omega$ is its angular velocity, measured via pulsar timing). Why? Because Jim's formalism shows (though Jim did not notice it) that a neutron star's moment of inertia is far more sensitive to the neutron star EOS than are the star's masses and radii; and astronomers expect near future observations of the binary pulsar J0737-3039 to reveal its moment of inertia to far better accuracy than any now known.[22]

Also especially powerful for EOS extraction has been the neutron-star tidal deformability $\Lambda = Q/\mathcal{E}$, where $\mathcal{E}$ is a tidal field applied to the neutron star by the gravity of some external body, and $Q$ is the quadrupole moment induced in the neutron star by that tidal field. Jim did not give a formula for $\Lambda$, but his formalism has been the foundation for deducing one. This deformability, like the moment of inertia, is highly sensitive to the EOS. The LIGO team has measured $\Lambda$ with moderately good accuracy using the observed gravitational waves from the binary-neutron-star merger GW170817, and has concluded that at (1-10)x $\rho_n$ the EOS is somewhat softer (lower dp/dρ) than some nuclear physics calculations have suggested.[23,24]

### EVOLUTION OF A BLACK HOLE INTERACTING WITH ITS ENVIRONMENT

In relativistic astrophysics, besides slowly rotating relativistic stars, Jim's other major contribution was his work with Stephen Hawking on the evolution of a black hole interacting with its environment. Jim first met Hawking at the Fifth Texas Symposium on Relativistic Astrophysics in Austin in 1970. The following year, Jim spent six months at the Institute for Theoretical Astronomy at Cambridge University working with Hawking on classical black holes, whose theory was rapidly developing at that time. This long collaboration would produce some of Jim's best work, including four joint papers, two of which brought about major changes in the field.

In their first paper,[25] they studied an old exact solution of the Einstein-Maxwell field equations found by Sudhansu Majumdar and Achilles Papapetrou in 1947.[26,27] Even though the field equations are nonlinear, Majumdar and Papapetrou showed that when the metric and Maxwell field are written in a certain way, the entire set of field equations reduce to a simple Laplace equation. The solution with point-like sources was thought to represent some kind of multi-particle system. Hartle and Hawking showed instead that this solution represents multiple charged black holes in static equilibrium, with their gravitational attraction exactly balanced by their electromagnetic repulsion. The locations of the point-like sources were actually finite area horizons and the solution could be extended inside. Although this first paper had no direct astrophysical relevance, it has been important because it was the first example of something that was often seen later in the development of supergravity: one can superpose supersymmetric black holes (and "black branes").

In their second paper, Hartle and Hawking developed a formalism to compute the flow of energy and angular momentum into a black hole's horizon, carried by matter fields and by gravitational perturbations.[28] Using this, they showed that there can be a significant coupling between the rotation of a black hole and a planet orbiting around it. In a follow-on paper by himself,[29] Jim computed the details of how the orbiting planet's tidal gravity deforms the horizon of a nonspinning black hole and discovered a remarkable feature: Unlike the outward deformation of the Earth produced by the Moon's gravity, which *lags* the Moon's location in the Earth's sky due to dissipation, the outward deformation of the hole's horizon *leads* the planet's location. The reason is simple but profound: The horizon is defined as the boundary of the hole's exterior from which photons can escape the hole's gravity, and its interior where they cannot. This definition is teleological — it depends on a future boundary condition rather than a past boundary condition.

Jim did not notice it, but his 1982 paper with Robert Geroch[30] contained the important insight that when a static tidal field is applied to a nonspinning black hole, there is no induced quadrupole moment. Thus, the tidal deformability $\Lambda = Q/\mathcal{E}$ of a nonspinning black hole vanishes, in contrast with that of a neutron star—a difference that might in the future become a powerful observational key for distinguishing black holes from neutron stars.

Jim's various analyses of the tidal coupling of black holes to their environments have major consequences today for the evolution of black holes in binary systems and the gravitational waves observed from such binaries. Particularly important for this has been Jim's 1985 paper with Thorne on the laws of motion and precession for black holes whose (intrinsic; not induced) multipole moments couple to external tidal fields.[31] For example, Hartle and Thorne deduced the details of how the coupling of an external (quadrupolar) tidal field to the hole's mass quadrupole moment, and the coupling of an external frame-drag field to the hole's current quadrupole moment, cause the hole's spin to precess, which shows up in a modulation of a binary's gravitational waveforms. Thorne, recalling this, his last collaboration with Jim, which entailed some very long and complex analytical calculations, says: "Jim taught me the power of wise laziness: Think through each step without doing it, and identify the minimum set of calculations that really have to be done before embarking on them."

## Quantum Properties of Black Holes

In 1974, Hawking discovered that black holes radiate with an essentially thermal spectrum.[32] This came as a surprise to everyone (including Hawking), and it took a while to be accepted. One questionable point in his derivation was the fact that the radiation seemed to come from modes near the horizon with frequency much greater than the Planck scale. Hawking's intuitive picture was that the radiation was a result of vacuum fluctuations that give rise to a pair of virtual particles, one with positive energy and one with negative energy. Near a black hole, the negative energy particle could fall behind the horizon and the positive energy particle could escape to infinity. Because negative energy particles can be viewed as positive energy particles moving backward in time, Hawking wanted another derivation of black hole radiation based on path integrals. As Hawking explained at Jim's sixtieth birthday conference in 1999:

> I wanted a mathematical treatment of black hole radiation as low energy particles leaking out of the horizon at late times, rather than as a high energy process during the collapse. I therefore wanted to use path integrals to calculate the amplitude for a particle to propagate from the future singularity of the black hole to an observer at infinity. All I knew about path integrals was the book by Feynman and Hibbs, which dealt only with the nonrelativistic case, and only in flat space. But Jim showed me how one could use path integrals to calculate the propagation of scalar particles in curved spacetime.

To make the path integral converge, Hartle and Hawking analytically continued some of the coordinates of the black hole metric. In the end, they showed that the probability that a particle escapes from the black hole is directly related to the probability that one fell in. In other words, black holes can be in thermal equilibrium. Furthermore, their new derivation made no reference to frequencies above the Planck scale. Their 1976 paper[33] was enormously influential for several reasons. It introduced a quantum state for matter outside a black hole, now known as the Hartle-Hawking vacuum, that has played a fundamental role in discussions of black hole thermodynamics ever since. In addition, their paper had another long-lasting legacy. As Hawking explained:

> Even more important than these results, this paper was the first to use Euclidean methods. We showed that the Schwarzschild solution could be analytically continued to a section on which it was Euclidean, that is to say, on which it had a positive definite metric. The natural choice of propagator was then the unique Green function on this Euclidean section. When one analytically continued this propagator back to the Lorentzian Schwarzschild solution, it had poles periodically in the imaginary time coordinate. This puzzled us, but Gary Gibbons and Malcolm Perry recognized them as the characteristic signature of thermal Green functions. This meant one could extend the proof of thermal emission, to interacting field theories as well.

Since then, Euclidean methods have become an indispensable part of quantum gravity research, including calculating nonperturbative processes, such as the pair creation of charged black holes in a background electric or magnetic field,[34] and the decay of certain higher dimensional (Kaluza-Klein) vacua by nucleating "bubbles of nothing."[35]

## Quantum Cosmology

In the late 1970s, Jim's interests turned to cosmology. Following the discovery that quantum particles are created in the early universe, Jim set out to calculate the backreaction of the created particles on the dynamics of the universe. In 1979 and 1980, he wrote influential papers with his postdoc Bei-Lok Hu and grad student Massimo Vincenzo Fischetti in which they developed a formalism to calculate this and applied it to simple models.[36,37,38]

The next step was to apply quantum mechanics to the spacetime itself and not just the matter propagating on it. This led to Jim's fourth, and most important, project with Hawking. It began at a Nuffield Workshop on the Very Early Universe that Hawking organized in Cambridge in June 1982. That workshop was dominated by discussions of the recently proposed theory of inflation. Some people said that inflation erased all traces of the initial conditions, so one did not have to understand how the universe began. But Hartle and Hawking understood that this could not be true, because one can take an arbitrarily inhomogeneous state today and evolve it back through any finite epoch of inflation and obtain some initial conditions near the Big Bang that will evolve to it. So inflation did not explain the observed homogeneity of the universe without some understanding of the initial state.

Hartle and Hawking knew that in simple (nongravitational) systems, one can obtain the ground-state wave function by a Euclidean path integral over all paths that start in the distant past at the classical ground state. In general relativity, this would correspond to a path integral over all positive definite four-dimensional metrics that approach flat Euclidean space asymptotically. This seemed an appropriate state to discuss the scattering of particles in asymptotically flat spacetimes, but it did not seem suitable for cosmology. During a visit by Hawking to Santa Barbara in the summer of 1982, he and Jim realized that there was another natural possibility: one can integrate over four-geometries that have no boundary in the past. In other words, the geometries in the path integral were analogous to a hemisphere, where the equator described a spatial three-geometry (i.e., a configuration of space), and the result of the path integral was the value of the wave function for this three-geometry. This can be described by saying "the boundary condition of the universe is that it has no boundary." Their 1983 paper created quite a sensation and started a wave of interest in quantum cosmology.[39] Their proposal quickly became known as the Hartle-Hawking no-boundary wave function of the universe.

Jim viewed this idea as his most important contribution to physics and spent much of the rest of his career studying its consequences in increasingly detailed models. When possible, he also compared its predictions with cosmological observations to check whether it indeed described our universe. As Jim later explained:

> It's been said that the signature of a great problem is one that leads to further great problems. It would I think be difficult to find a clearer example of this than the search for the quantum state of the universe and the understanding of the no-boundary wave-function of the universe. Indeed it would not be an exaggeration to say that I have spent a large part of my modest career working on the no-boundary wave-function and the great problems it has led to, confident that it will lead to many more because of the vast sweep and scope of phenomena that it can be applied to in a manageable way.[40]

Jim liked to emphasize that the no-boundary wave function shows that the usual dichotomy in physics between dynamics and initial conditions can be transcended. The path integral involves the action that determines the dynamics. But in the no-boundary wavefunction, it also determines the initial state. No other initial conditions are needed.

In addition to studying the consequences of his no-boundary wave function, Jim thought deeply about what it means to apply quantum mechanics to the universe as a whole. It was clear that some modification of the standard approach to quantum mechanics was needed, because there were no outside observers who could do measurements on the system. (Jim's interest in applying quantum mechanics to individual systems started early in his career, and he published an important paper in 1968 on this topic.[41]) In a long-term collaboration with Gell-Mann, Jim worked to develop a formulation of quantum mechanics, known as the consistent histories formulation, that was sufficiently general to describe closed systems like the universe. He explored fundamental questions such as: how does the current semiclassical epoch emerge from a quantum universe, and how do we describe a universe with a final boundary condition as well as an initial boundary condition?

Hartle and Gell-Mann wrote a total of nine papers together (not including two papers that Jim added Gell-Mann's name to after he died in 2019). Jim thought that to a good approximation Gell-Mann knew everything. Conversations with him were not confined to physics, and he felt they were a true adventure of the mind.

Jim became a frequent visitor to the Santa Fe Institute after Gell-Mann moved there from Caltech in 1993. This led to him being appointed a Santa Fe Institute External Faculty Member in 2006.

Jim thought deeply about many of the puzzles of quantum mechanics and wrote short essays explaining his views. They included such titles as "Living in a Superposition," "Quantum pasts and the Utility of History," "Why Our Universe is Comprehensible," and "Quantum physics and Human Language." Many of them were published in 2021 in a volume entitled *The Quantum Universe*.[42]

Jim's scientific contributions were recognized by many honors, including election as a Fellow of the American Academy of Arts and Sciences in 1989 and member of the National Academy of Sciences in 1991. In 2009, he won the Einstein Prize from the American Physical Society.

## THE INSTITUTE FOR THEORETICAL PHYSICS

In 1977, the NSF solicited proposals for a new kind of physics institute where people from all over could gather for an extended stay and work on problems in theoretical physics. Jim and three of his colleagues at UCSB (Ray Sawyer, Bob Sugar, and Doug Scalapino) decided to submit a proposal. The competition was intense, and only one proposal was going to be funded. To their delight, the NSF chose UCSB. The next step was to find a director, and Jim and his colleagues persuaded Walter Kohn to move from San Diego to Santa Barbara. In fall 1979, the Institute for Theoretical Physics (ITP) began operation with a long program on quantum gravity, with Jim playing an active role. The initial ITP grant was for five years, and there was concern that ITP might shut down at the end of this period, but it was renewed for another five years and has been renewed ever since.

Walter Kohn was succeeded by Bob Schieffer and then Jim Langer as director. In 1995, the ITP was looking for another director and asked Jim to serve. Although he was not eager to take on this role, he felt obliged to help the organization he co-founded and agreed to serve as its "interim" fourth director. It took Jim only two years to recruit David Gross from Princeton, and in 1997, Jim happily returned to the physics department and resumed his teaching and research.

Gross started fund-raising, and with a major donation from the Kavli Foundation, the ITP became the Kavli Institute for Theoretical Physics (KITP) in 2002. It has now been in operation for more than forty-five years and has been tremendously successful, with about 1,000 physicists a year participating in its programs and conferences.

## LATER CAREER AND RECOLLECTIONS

In the 1980s, Jim started to teach an undergraduate course on general relativity. Rather than follow the traditional approach of first explaining the advanced mathematics (differential geometry) required for Einstein's equation, Jim decided to try something new. He found that he could explain how to extract physical predictions from a curved spacetime geometry relatively easily, without most of the complicated machinery of differential geometry. He then wrote down the curved geometries that describe black holes and Big Bang cosmology and derived most of the important predictions of general relativity. Only at the end, if there was time, would Jim explain where these curved spacetimes come from, as solutions to Einstein's equation. Jim called this a "physics first" approach to general relativity, and it has become very popular. Over the years, Jim refined his lecture notes, and in 2003 he published them as a book entitled *Gravity: An Introduction to Einstein's General Relativity*.[43] Many people saw the value of Jim's new approach and adopted it for their own courses. It is currently used at about one hundred universities.

In the last few years of his life, Jim composed several essays recalling his experience working with many prominent physicists, including Stephen Hawking,[44] John Wheeler,[45] and Nobel laureates Richard Feynman,[46] Kip Thorne,[47] and Steven Weinberg.[48] These memoirs, which he published on the open-access arXiv platform, provide first-hand insights into how theoretical physics progresses.

Jim was a modest man and often minimized his own achievements. When talking about the Hartle-Hawking no-boundary wave function, he often called it "Hawking's wave function." He later admitted that he did so to strengthen the case for a Nobel Prize for Hawking.[49] Jim also had a wonderful sense of humor that made it a genuine pleasure to be his colleague. For example, Jim once reflected on the fact that his two most famous contributions were his Hartle-Hawking vacuum state and Hartle-Hawking no-boundary wave function and quipped, "Why couldn't I be known for something rather than nothing!"

Jim loved to make up toasts for various occasions which were often clever and humorous. Two of his favorites were: "I would like to toast gravity—it keeps our feet on the ground, and for some of us, puts money in our pockets," and, "The famous American writer Mark Twain said that the road to good health consists of eating what you don't want, drinking what you don't like, and doing what you'd rather not. I have to tell you that this evening, we are not on the road to good health."

Jim also wrote the following theme song for relativity, which he sang with great gusto at all possible occasions:

> Oh Relativity 'tis to thee
> That we pledge our loyalty
> Strings may come and strings may go
> But there is one thing that we know

Einstein said spacetime was right
We won't give up without a fight
O relativity tis to thee
That we pledge our loyalty

Not only was Jim a brilliant physicist, he was a remarkably kind and caring person. He was genuinely interested in the students and postdocs in his group and met with them frequently to see how they were doing and offer advice. Jim stopped taking Ph.D. students in 1998 because his students were doing quantum cosmology and having a hard time finding postdoc positions in a research environment dominated by string theory and loop quantum gravity. In total, Jim supervised thirteen Ph.D. students during his career, and he kept in touch with many of them and offered support when needed. As former student David Craig said, "Jim remained a steadfastly supportive mentor and friend to the end. His support was crucial to enabling me to remain in science more than once, and for that I will be forever grateful." Former student Paul Anderson noted, "Unknown to me until decades later, Jim played a major role in helping me to stay in academia during my postdoc years."

Jim also cared about the broader gravitational physics community. For many years, he had annual phone calls with the head of the gravitational physics division at the National Science Foundation about whether the best young people were finding postdoc positions. He then kept track of them to see if they were getting faculty jobs.

After Jim passed away on May 17, 2023, his family set up a website where people could post comments about him. Within a couple weeks, it was filled with dozens of testimonials describing all the ways that Jim helped these people in their careers. There were also many comments about Jim's impact on the gravitational physics culture. Here are a few examples.

Don Marolf wrote:

Jim always emphasized the importance of looking after others, being responsible, and creating a positive supportive environment for young scientists… I came to see Jim as a primary reason why the U.S. gravitational physics community had so often seemed like a warm and welcoming family.

Neil Turok wrote:

Jim was quite unique, in my experience, in truly seeing the big picture—beyond technicalities, beyond personalities or status, beyond current fashions. It is an extraordinary privilege to be able to think about the universe, its origin and destiny, with at least some aspiration to scientific rigor. It is an even greater privilege to share the wonder and joy we find in that activity with other people. No-one in the field set a better example of personal humility, unselfish dedication and support for young scientists than Jim.

## Selected Bibliography